\documentclass{llncs}
\usepackage{llncsdoc}
\usepackage{bbm}
\usepackage{amsmath}
\usepackage{amsfonts}
\usepackage{mathrsfs}
\usepackage{amssymb}
\usepackage{graphicx}
\usepackage{verbatim}
\usepackage[ruled,vlined,]{algorithm2e}
\usepackage{array}
\usepackage[all]{xy}
\usepackage{authblk}
\usepackage{sidecap}
\usepackage{multirow}
\usepackage{eso-pic}

\newtheorem{df1}{Definition}
\newtheorem{th1}[df1]{Theorem}
\newtheorem{le1}[df1]{Lemma}

\newtheorem{pr1}[df1]{Proposition}

\newcommand{\utl}{\,\mathcal{U}\,}
\newcommand{\mc}{\mathcal{M}}
\newcommand{\pmc}{\mathcal{M}_\ast}

\title{Asymptotic Bounds for Quantitative Verification of Perturbed Probabilistic Systems\thanks{The work is supported by grant R-252-000-458-133 from Singapore Ministry of Education Academic Research Fund. The authors would like to thank Professor Mingsheng Ying for pointing them to perturbation theory and the anonymous referees for improving the draft of this paper.}}
\author{Guoxin Su  \and David S.~Rosenblum}
\institute{
National University of Singapore\\
\email{$\{$sugx,\,david$\}$@comp.nus.edu.sg}}

\begin{document}
\maketitle

\AddToShipoutPicture*{\put(146,740){A short version of the paper is published in the proceedings of ICFEM'13}}

\begin{abstract}
The majority of existing probabilistic model checking case \mbox{studies} are based on well understood theoretical models and distributions. However, real-life probabilistic systems usually involve distribution parameters whose values are obtained by empirical measurements and thus are subject to small perturbations. In this paper, we consider perturbation analysis of reachability in the parametric models of these systems (i.e., parametric Markov \mbox{chains}) equipped with the norm of absolute distance.
Our main contribution is a method to compute the \emph{asympto\-tic \mbox{bounds}} in the form of \emph{condition numbers} for constrained reachability probabilities against perturbations of the distribution parameters of the system. The adequacy of the method is demonstrated through experiments with the Zeroconf protocol and the hopping frog problem.
\end{abstract}

\section{Introduction}

Probabilistic model checking is a verification technique that has matured over the past decade, and one of the most widely known and used probabilistic model checking tools is PRISM \cite{kwiatkowskaetal11}.
The majority of the reported case studies of probabilistic model checking, including those performed in PRISM, involve systems whose stochastic nature arises from well understood theoretical probabilistic distributions, such as the use of a fair coin toss to introduce randomization into an algorithm, or the uniform distribution of randomly chosen IP addresses in the Zeroconf protocol. More complex, realistic systems, on the other hand, involve behaviors or other system characteristics generated by empirical distributions that must be encoded via empirically observed parameters. In many cases, these distribution parameters are based on finite numbers of samples and are statistical estimations that are subject to further adjustments. Also, the stochastic nature of the model (e.g., the failure rate of some hardware component) may be varying over time (e.g., the age of the component). The conventional techniques and tools of probabilistic model checking, including PRISM, do not provide sufficient account for systems with distribution parameters. Consider, for instance, the setting of automata-based model checking: Given a (probabilistic) model $\mc$ and an LTL formula $\varphi$, the model checker returns a satisfaction probability $p$ of $\varphi$ in $\mc$. However, $\mc$ is just an \emph{idealized} model of the probabilistic system under consideration, and because the real distribution(s) of its parameter(s) may be slightly different from those specified in $\mc$, $p$ is merely a \emph{referential} result whilst the \emph{actual} result may deviate from $p$ to a small but non-trivial extent. We elaborate this pitfall in the following two concrete examples.

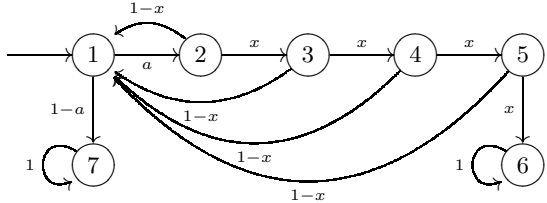
\begin{figure}[t]
$$
\entrymodifiers={++[o][F-]}
\SelectTips{cm}{}
\xymatrix  {
*\txt{} \ar[r]
& 1  \ar[r]_a \ar[d]_{1-a}
& 2 \ar[r]^x\ar@/_1pc/_{1-x}[l]
& 3 \ar[r]^x
\ar@/^1.5pc/ ^{1-x}[ll]
& 4 \ar[r]^x
\ar@/^2.8pc/ ^{1-x}[lll]
& 5 \ar@/^4pc/ ^{1-x}[llll] \ar[d]_{x}\\
*\txt{} & 7 \ar@(ul,dl)[]_1& *\txt{} & *\txt{} &*\txt{} & 6 \ar@(ul,dl)[]_1 }
$$
\caption{Zeroconf protocol with noisy channels}\label{fig:zeroconf}
\end{figure}

\paragraph{Motivating examples.} We first consider an IPv4 Zeroconf protocol model for a network with noisy communication channels. Figure \ref{fig:zeroconf} presents the protocol model that uses a maximum of four ``ok'' message probes. Let $a$ be the probability that the new host chooses an IP address that has been assigned already, and $x$ be the probability that the probe or its reply is lost due to channel noise or some other reason (if any). If an IP address is randomly chosen, then $a$ is equal to $m/n$, where $n=60,534$ is the size of IP address space as specified in Zeroconf and $m$ is number of hosts already connected. By contrast, $x$ relies on an \emph{ad hoc} statistical estimation of the loss rate of messages tested in experiments. In reality, it is less meaningful to specify a single, constant value of $x$, as the measurement could be affected by factors such as network load, environment noise, temperature, measurement time, etc. Instead, $x$ may be given as the expression $x_0\pm  x_\Delta$, where $x_0$ is the mean value of the measured results and $x_\Delta$ specifies the maximal perturbation. It is therefore reasonable to express the probability that an address collision happens as $p=p_0\pm y_\Delta$, where $p_0$ is a referential value for the result and $y_\Delta$ specifies the range of perturbation of $p$.
However, although the standard model checking techniques allow one to obtain $p_0$ by inputting $x_0$, they provide little account for the relationship between $y_\Delta$ and $x_\Delta$.

Another example is a variant of the classic hopping frog problem. A frog hopping between four rocks and the $(i,j)$-entry in the following \emph{parametric transition matrix} provides the concrete or abstract probability of frog's movement from the $i$th rock to the $j$th rock:
\begin{align*}
    \left(
          \begin{array}{cccc}
            ~z_1~ & ~z_2~ & ~z_3~ \vspace{0.2em} & ~z_4~\\
            \frac{3}{8} & \frac{1}{8} & \frac{1}{4} & \frac{1}{4}\vspace{0.2em}\\
            0 & \frac{1}{2} & \frac{1}{2} & 0\vspace{0.2em}\\
            \frac{1}{3} & 0 & \frac{1}{3} & \frac{1}{3}\vspace{0.2em}\\
          \end{array}
        \right)
\end{align*}
where the tuple of variables $(z_1,z_2, z_3,z_4)$ satisfies that $z_i\geq 0$ for each $1\leq i\leq 4$, $z_1+ z_2+z_3+z_4=1$ and
\begin{align}\label{eq:froghopping}
    \left|z_1- \frac{3}{8}\right| + \left|z_2-\frac{1}{8}\right| + \left|z_3-\frac{1}{4}\right| + \left|z_4-\frac{1}{4}\right|\leq \Delta
\end{align}
with $\Delta$ a sufficiently small positive number. Intuitively, according to  Equation \eqref{eq:froghopping}, $(\frac{3}{8}, \frac{1}{8},\frac{1}{4},\frac{1}{4})$ is the idealized distribution  of $(z_1,z_2,z_3,z_4)$ and a small perturbation of $(z_1,z_2,z_3,z_4)$ is allowed and measured.
We call $(z_1,z_2, z_3,z_4)$ a \emph{distribution parameter} and $(\frac{3}{8}, \frac{1}{8},\frac{1}{4},\frac{1}{4})$ its \emph{reference}. A typical model checking problem for this example can be stated as ``what is the probability that the frog eventually reaches the fourth rock without landing on the third rock?''
Again, well established model checking techniques do not provide a direct solution for this question.

\paragraph{Approach.}

In a nutshell, the aforementioned two examples demonstrate that a satisfactory model checking result for a probabilistic system with one or more perturbed distribution parameters should shed light on the \emph{sensitivity} of the result to the distribution parameters. In this paper, we provide a method to compute the \emph{asymptotic bounds} of the results in terms of \emph{condition numbers} for reachability checking of probabilistic systems under perturbations. We model the probabilistic systems in discrete-time Markov chains (MCs)\footnote{Throughout the paper, unless mentioned otherwise, by MCs we mean discrete-time Markov chains.} with distribution parameters, which are coined as \emph{parametric Markov chains} (PMCs), and introduce the \emph{norm of absolute distance} to measure the deviation distances of their distribution parameters (as exemplified by equation \eqref{eq:froghopping}). The reachability checking is formalized as follows: Given a PMC $\pmc$ with state space $S_{\pmc}$ and two sets of states $S_?, S_!\subseteq S_{\pmc}$, a reachability problem is phrased as the probability of ``reaching \mbox{states} in $S_!$ only via states in $S_?$''. By adopting a notation from temporal logic, the problem is denoted by $S_?\utl S_!$, where $\utl$ refers to the ``until'' operator.\footnote{In fact, the formulation of reachability in the present paper is sightly more general than the standard definition of reachability and sometimes is called \emph{constrained reachability}, since the $S_?$ in $S_?\utl S_!$ plays a constraining role.} Two instances of the reachability problem class are mentioned in the two motivating examples above.
The output of the reachability checking contains a referential probabilistic result $p\in  [0,1]$ and a condition number $\kappa_i\in \mathbb{R}$ where $i\in I$, an index set, for \emph{each} distribution parameter. The significance of the output is that, if a sufficiently small perturbation $\Delta_i$, measured by the \mbox{norm} of absolute distance, occurs on the parameter whose condition number is $\kappa_i$ for each $i\in I$, then the actual result is asymptotically bounded by $p\pm \sum_{i\in I}\kappa_i\Delta_i$.
A brief comparison of the reachability checking in MCs and PMCs in terms of input and output is presented in Table \ref{tb:mcproblem}.

\begin{table}[t]
\caption{Reachability checking in MCs and PMCs} \label{tb:mcproblem}
\centering
\begin{tabular}{c c c }
  \hline\noalign{\smallskip}
  ~Model~ & ~Input~ & ~Output~  \\ \noalign{\smallskip}
   \hline\noalign{\smallskip}
   ~$\mc$~ & ~$S_?\utl S_!$~ & $p$ (idealized reachability probability)~ \vspace{.25em}\\
 \multirow{2}{*}{~$\pmc$} & \multirow{2}{*}{~$S_?\utl S_!$~}  & ~$p$ (referential reachability probability) \vspace{-.25em} \\
 && ~and $\kappa_i$ (condition numbers) \\
  \noalign{\smallskip} \hline
\end{tabular}
\end{table}

Perturbation bounds have be pursued for MCs in a line of research \cite{schweitzer68,chomeyer00,solanvieille03,heigergott08}. However, to the best of our knowledge, this paper is the first one devoted to the application of \mbox{concepts} and methods from perturbation theory to quantitative verification.
To further explain our method, it is beneficial to compare it with a standard method for error estimation based on differentiation and linear approximation.
Suppose a sphere (such as a prototype of balls produced by a sporting goods factory) is measured and its radius is $21cm$ with a possible small error within $0.05cm$. The dependence of the sphere volume on the radius is given by $V = \frac{4}{3}\pi r^3$.
The problem is to compute volume error $V_\Delta$ given the radius error $r_\Delta$. We recall a classic method for this problem: First, the differential function of $V$ on $r$ is given by $\mathrm{d}V = 4\pi r^2 \mathrm{d}r $.
Second, let $\mathrm{d}r =r_\Delta=0.05cm$ (which is significantly small compared with $r=21cm$) and we obtain $V_\Delta\approx \mathrm{d}V=4\pi\times 21^2\times 0.05 \approx 277 cm^3$. The sensitivity of $V_\Delta$ to $r_\Delta$ is approximated by the ratio
$\frac{\mathrm{d}V}{\mathrm{d}r}= 4\pi r^2 \approx 5,542$
and this expression is useful if the value of $r_\Delta$ is unknown in advance.
We aim to develop a similar methodology to estimate the perturbations of reachability in PMCs, which is comparable to the use of differentiation and linear approximation in estimating the error of the ball volume described above.

\paragraph{Organization.} The remainder of the paper is organized as follows. The next section (Section \ref{sec:pmc}) presents the formulations of PMCs and introduces the norm of absolute distance for probabilistic distributions. For presentation purposes, we separate the treatment of PMCs into that of basic PMCs, which have a single distribution parameter, and general PMCs, which have multiple distribution parameters. Section \ref{sec:ana-bpmc} deals with basic PMCs by establishing a method for computing their asymptotic bounds, in particular, condition numbers for the given reachability problems. Section \ref{sec:ana-pmc} generalizes the computation method to handle non-basic PMCs. Our approach is evaluated by experiments in Section \ref{sec:experiment}. Related work is discussed in Section \ref{sec:relatedwork}. In Section \ref{sec:conclu}, the paper is concluded and several future research directions are outlined.
Proof details of the theorems are found in the Appendix.

\section{Parametric Markov Chains}\label{sec:pmc}

In this section, we define the formal models of PMCs, which are parametric variants of MCs. Informally speaking, a PMC is obtained from an MC by replacing the \emph{non-zero} entries in one or more rows of its probabilistic transition matrix by mutually distinct variables.

Let $\mathbf{x}$ be a symbolic vector of pair-wise distinct variables, called a \emph{vector variable} for short. A \emph{reference} $\mathbf{r}$ for $\mathbf{x}$ is a probabilistic vector such that $|\mathbf{r}|=|\mathbf{x}|$. We use $\mathbf{x}[j]$ to denote the variable in the $i$th entry of $\mathbf{x}$. An extension of $\mathbf{x}$, denoted by $\mathbf{x}^\ast$, is obtained by inserting the number zero into $\mathbf{x}$, {i.e.,}
\begin{align*}
    \mathbf{x}^\ast= ( \underbrace{0,\ldots, 0}_{l_0 \, 0\hbox{'s}},\mathbf{x}[1],\ldots,\mathbf{x}[j],\underbrace{0,\ldots, 0}_{l_j \, 0\hbox{'s}},\mathbf{x}[j+1],\ldots, \mathbf{x}[k], \underbrace{0,\ldots, 0}_{l_k \, 0\hbox{'s}} ) \enspace,
\end{align*}
where $l_0,\ldots, l_k$ are non-negative integers. Two vector variables are \emph{disjoint} if they share no common variables. Let $(\mathbf{x}_i)_{i\in I}$ be a sequence of pair-wise disjoint vector variables for an index set $I\neq\varnothing$ of positive integers. We abbreviate the sequence $(\mathbf{x}_i)_{i\in I}$ as $\mathbf{x}_I$. Let $\mathcal{P}(\mathbf{x}_I)$ be a $k\times k$ abstract square matrix with parameters $\mathbf{x}_I$ such that (i) $k\geq \max(I)$, (ii) if $i\notin I$ then the $i$th row of $\mathcal{P}(\mathbf{x}_I)$ is a probabilistic vector and (iii) if $i\in I$ then the $i$th row of $\mathcal{P}(\mathbf{x}_I)$ is $\mathbf{x}_i^\ast$, an extension of $\mathbf{x}_i$. Here, the involvement of extensions of vector variable intends to be consistent with the replacement of non-zero entries by variables mentioned previously.  Such an abstract matrix $\mathcal{P}(\mathbf{x}_I)$ is called a \emph{parametric transition matrix} and each parameter $\mathbf{x}_i$ appearing in $\mathcal{P}(\mathbf{x}_I)$ is called a \emph{distribution parameter}. We can view $\mathcal{P}(\mathbf{x}_I)$ as mapping from sequences of vectors to concrete matrices. As such, $\mathcal{P}\langle \mathbf{r}_I\rangle$, where $\mathbf{r}_I$ abbreviates $(\mathbf{r})_{i\in I}$, is the matrix obtained by replacing $\mathbf{x}_i$ with its reference $\mathbf{r}_i$ for each $i\in I$. Sometimes, especially in our running examples, it is cumbersome to present the distribution parameters $\mathbf{x}_I$ in $\mathcal{P}(\mathbf{x}_I)$; if so, we just write $\mathcal{P}$ and mention its distribution parameters in the text.

\begin{df1}\label{df:pmc}
A \emph{parametric Markov chain} (PMC) is given by the tuple
\begin{align*}
    \pmc=(\iota, \mathcal{P}( \mathbf{x}_I), \mathbf{r}_I)\enspace,
\end{align*}
where $\iota$ is a probabilistic vector (for the initial distribution of $\pmc$), $\mathcal{P}( \mathbf{x}_I)$ is a $|\iota|\times |\iota|$ parametric transition matrix, and $\mathbf{r}_I$ contains references for vector variables in $\mathbf{x}_I$.
\end{df1}

The underlying MC of $\pmc$ is $\mc=(\iota, \mathcal{P}\langle \mathbf{r}_I\rangle)$. We do not specify the \mbox{state} space for $\pmc$ and $\mc$. But throughout the paper, we assume that their state spaces $S_{\pmc}=S_{\mc}=\{1, \ldots, |\iota|\}$.

As promised earlier, we introduce a statistical distance measurement between distribution parameters and their references, which is given by the norm of absolute distance (also called total variation).
\begin{df1}
The statistical distance for $\pmc$ is given by $\|\cdot \|$ such that $\|\mathbf{v}\|=\sum_{i=1}^n |\mathbf{v}[i]|$ for any vector $\mathbf{v}$.
\end{df1}

By definition, the scalar function $\|\mathbf{x}^\ast-\mathbf{r}^\ast\|$ is the same as the scalar function $\|\mathbf{x}-\mathbf{r}\|$. If $I$ is a singleton, we also call the PMC a \emph{basic PMC}. In other words, a basic PMC is a PMC with a single distribution parameter.

We now present examples of PMCs. The first example is a PMC $\pmc^\mathrm{fg}$ for the hopping frog. Its parametric transition matrix (the $4\times 4$ symbolic matrix already presented in the Introduction) is denoted by $\mathcal{P}_\ast^\mathrm{fg}$. In $\mathcal{P}_\ast^\mathrm{fg}$, the tuple $(z_1,z_2,z_3,z_4)$ is given as the only distribution parameter. We let the reference to the parameter be $(0.375,0.125,0.25,0.25)$. In words, ideally, the probabilities for the frog to jump from the first rock to the first, second, third, and fourth rocks are $0.375$, $0.125$, $0.25$, and $0.25$, respectively. Such a PMC is denoted by $\pmc^\mathrm{fg}$. Additionally, we let the initial distribution in $\pmc^\mathrm{fg}$ be $\iota^{\mathrm{fg}}=(0.25,0.25,0.25,0.25)$, which means that all rocks have an equal probability to be the frog's initial position. Clearly, $\pmc^\mathrm{fg}$ is a basic PMC.

Another example is PMC $\pmc^\mathrm{zf}$ for the noisy version of Zeroconf. For illustration purposes, a probabilistic transition system with a parameter $x$ is provided in Figure \ref{fig:zeroconf}. The formulation of $\pmc^\mathrm{zf}$ according to Definition \ref{df:pmc} deviates from the transition system because of the use of distribution variables. Following the definition, we let the sequence of distribution parameters be $(x_i,x_i')_{i =1}^4$. The parametric transition matrix is given by the following $7\times 7$ symbolic matrix:
\begin{align*}
   \mathcal{P}^{\mathrm{zf}}= \left(
          \begin{array}{ccccccc}
            0     &~a~&~0~  &~0~  &~0~  &~0~  &~1-a\\
           ~x_1~& 0 & x_1' & 0   & 0   & 0   & 0 \\
            x_2 & 0 & 0   & x_2' & 0   & 0   & 0  \\
            x_3 & 0 & 0   & 0   & x_3' & 0   & 0  \\
            x_4 & 0 & 0   & 0   & 0   & x_4' & 0  \\
            0     & 0 & 0   & 0   & 0   & 1   & 0  \\
            0     & 0 & 0   & 0   & 0   & 0   & 1  \\
          \end{array}
        \right)
\end{align*}
The constant number $a$ is calculated according to the number of addresses and that of the occupied ones. The reference for $(x_i,x_i')$ in $\pmc^\mathrm{zf}$ is $(0.75, 0.25)$ for each $1\leq i\leq 4$. In other words, we suppose that under idealized conditions the chances of not receiving a reply in four probes are equivalently $0.25$. The initial distribution $\iota^\mathrm{zf}$ is $(1, 0,\ldots,0)$, as state $1$ is the initial state.

\section{Perturbation Analysis of Basic PMCs}\label{sec:ana-bpmc}

From this section, we commence the perturbation analysis of reachability problems in PMCs. For presentation purposes, in this section we deal with basic PMCs. Recall that a basic PMC has a single distribution parameter. Our main goal is to establish a method to compute an asymptotic bound, in particular, a condition number for a given reachability problem in a basic PMC against the perturbation of its sole distribution parameter. In the next section, we generalize the method to the setting of general PMCs.

\subsection{Perturbation Function}

Throughout this section, we assume $\pmc$ contain a single distribution parameter; thus, $\pmc=(\iota, \mathcal{P}(\mathbf{x}), \mathbf{r})$. Without loss of generality, let $\mathbf{x}$ appear in the first row of $\mathcal{P}(\mathbf{x})$. We consider the reachability problem $S_?\utl S_!$ in $\pmc$ with state space $S_{\pmc}=\{1, \ldots, |\iota|\}$ such that $S_?\cup S_!\subseteq S_{\pmc}$. For convenience, we let $S_?=\{1,\ldots, n_?\}$ and $S_!=\{n_!,\ldots, |\iota|\}$, where $0\leq n_?<n_!\leq |\iota|$. Thus, $S_?\cap S_!=\varnothing$.\footnote{This assumption does not impose any theoretical restriction on the reachability problem, because if $S_?\cap S_!\neq \varnothing$ then we carry out the analysis based on $(S_?\backslash S_!)\utl S_!$.} We call $S_?$ the \emph{constraint} set of $S_?\utl S_!$ and $S_!$ its \emph{destination} set. In the remainder of this subsection, our goal is to formulate a function that captures the effect of the perturbation of $\mathbf{x}$ on the probability of $S_?\utl S_!$ being satisfied by $\pmc$. To motivate and explain the formulation, we recall the standard model checking techniques for reachability probabilities based on non-parametric MCs.

The underlying MC of the basic PMC $\pmc$ is $\mc=(\iota, \mathcal{P}\langle \mathbf{r}\rangle)$ and the state space of $S_{\mc}=S_{\pmc}$. Let $\mathcal{P}'=\mathcal{P}\langle \mathbf{r}\rangle$. We use $\mathcal{P}'[i,j]$ to denote the number in the $(i,j)$-entry of $\mathcal{P}'$.
Let $\mathbf{p}$ be a vector such that $|\mathbf{p}|=n_?$ and, for each $1\leq i\leq n_?$, $\mathbf{p}[i]$ is the probability of $S_?\utl S_!$ satisfied in state $i$ of $\mc$. Thus,
\begin{align}\label{eq:reachprob}
\mathbf{p}[i]=\sum_{j=1}^{n_?}\mathcal{P}'[i,j]\cdot \mathbf{p}[j] + \sum_{j=n_!} ^{|\iota|} \mathcal{P}'[i,j]\enspace,
\end{align}
for each $1\leq i\leq n_?$.
We rewrite the equation system given in \eqref{eq:reachprob} as
\begin{align}\label{eq:matrixformula}
    \mathbf{p}=  \mathbf{A}'\cdot \mathbf{p}+ \mathbf{b}' \enspace,
\end{align}
where $\mathbf{A}'$ is the up-left $n?\times n?$ sub-matrix of $\mathcal{P}'$ (thus, $\mathbf{A}[i,j]=\mathcal{P}[i,j]$ for each $1\leq i,j\leq n_?$), and $\mathbf{b}'$ is a vector such that $|\mathbf{b}'|=n_?$ and $\mathbf{b}[i]=\sum_{j=n_!} ^{|\iota|} \mathcal{P}[i,j]$ for each $1\leq i\leq n_?$. Moreover, $\mathbf{p}$ is the least fixed point satisfying equation \eqref{eq:matrixformula}.

\begin{le1}\label{le:preachability}
$\mathbf{p}$ is computed by $\mathbf{p} = \sum_{i=0}^\infty {\mathbf{A}'}^i\cdot \mathbf{b}'$.
\end{le1}

In the following, we define the parametric counterparts of $\mathbf{A}'$ and $\mathbf{b}'$ specified in Equation \eqref{eq:matrixformula}, namely, $\mathbf{A}(\mathbf{x})$ and $\mathbf{b}(\mathbf{x})$. It should be stressed that according to our notations not necessarily all variable in the vector variable $\mathbf{x}$ appear in each of $\mathbf{A}(\mathbf{x})$ and $\mathbf(b)(\mathbf{x})$. There are two equivalent ways to obtain $\mathbf{A}(\mathbf{x})$ and $\mathbf{b}(\mathbf{x})$. One way is to define them by going over the aforementioned procedure for $\mathbf{A}'$ and $\mathbf{b}'$, and the other way is to directly parameterize $\mathbf{A}'$ and $\mathbf{b}'$. Here, the second way is chosen.
Recall that the first row of $\mathbf{P}(\mathbf{x})$ is an extension $\mathbf{x}^\ast$ of $\mathbf{x}$. We let $\mathbf{x}^\ast|_{n_?}$ be the sub-vector of $\mathbf{x}^\ast$ that consists of the first $n_?$ components (variables or zeros) of $\mathbf{x}^\ast$, and $\overrightarrow{\mathbf{x}}_{n_!}$ be the expression $\mathbf{x}[n_!]+\ldots+\mathbf{x}[|\iota|]$. Then, $\mathbf{A}(\mathbf{x})$ is obtained by replacing the first row in $\mathbf{A}'$ with $\mathbf{x}^\ast|_{n_?}$ and $\mathbf{b}(\mathbf{x})$ is by replacing the first entry of $\mathbf{b}$ with $\overrightarrow{\mathbf{x}}_{n_!}$.
If it is not necessary to mention the (possible) variables in $\mathbf{A}(\mathbf{x})$ or $\mathbf{b}(\mathbf{x})$, we just write $\mathbf{A}$ or $\mathbf{b}$.

As an example, consider the following model checking problem of the hopping frog (which has already been mentioned in Sections 1 and 2): What is the probability of reaching the fourth rock without landing on the third one? In this problem, the constraint set is $\{1,2\}$ and the destination set is $\{4\}$. Recall that the only distribution parameter in $\pmc^\mathrm{fg}$ is $(z_1,z_2,z_3,z_4)$. Thus, the parametric matrix and the parametric vector are respectively given by
\begin{align*}
\mathbf{A}^{\mathrm{fg}} =
\left(
         \begin{array}{cc}
    z_1 & z_2 \\
    \frac{3}{8} & \frac{1}{8} \\
         \end{array}
       \right)
\enspace, & \quad
\mathbf{b}^{\mathrm{fg}}=
\left(
         \begin{array}{cc}
    z_4 \\
    \frac{1}{4} \\
         \end{array}
       \right)\enspace.
\end{align*}

Let $\mathbf{V}= [0,1]^{k}$ where $k=|\mathbf{x}|$ and $\mathbf{U}=\{\mathbf{v}\in \mathbf{V} ~|~ \sum_{i=1}^n \mathbf{v}[i]=1\}$. Let $\iota_?$ be the first $n_?$ items in $\iota$.
\begin{df1}\label{df:rho}
The perturbation function of $\mathbf{x}$ for a basic PMC $\pmc=(\iota, \mathcal{P}(\mathbf{x}), \mathbf{r})$ and with respect to the problem $S_?\utl S_!$ such that $S_?, S_!\subseteq S_{\pmc}$ is $\rho:\mathbf{V}\rightarrow [-1,1]$ such that
\begin{align}\label{eq:rho}
    \rho(\mathbf{x})
    %= &~\sum_{i=1}^{n_?} \iota[i]\sum_{j=0}^\infty \left(\mathbf{A}(\mathbf{x})^j\cdot \mathbf{b}(\mathbf{x})-\mathbf{A}(\mathbf{r})^j\cdot \mathbf{b}\right)[i] \nonumber\\
    =& ~ \iota_{?}\cdot \sum_{j=0}^\infty \left(\mathbf{A}(\mathbf{x})^j\cdot \mathbf{b}(\mathbf{x})-\mathbf{A}\langle\mathbf{r}\rangle^j\cdot \mathbf{b}\langle\mathbf{r}\rangle\right)\enspace.
\end{align}
%\begin{align}\label{eq:rho}
   % \rho(\mathbf{x})= &~\sum_{i=1}^{n_?} \iota[i](f(\mathbf{x})[i]-f(\mathbf{r})[i]) \nonumber\\
   % =& ~ \iota_{?}\cdot (f(\mathbf{x})-f(\mathbf{r}))\enspace,
%\end{align}
%where $f(\mathbf{x})$ is given by
%\begin{align}\label{eq:ef}
   % f(\mathbf{x})=& ~ \sum_{j=0}^\infty \mathbf{A}_{\mathbf{x}^\ast_{?}}^j\cdot \mathbf{b}_{ \widehat{\mathbf{x}}^\ast_{!}}\enspace.
    %f(x)=& ~ \sum_{i=0}^\infty A[I:=x_{\leq m}]^i\cdot u[I:= \sum x_{\geq n}]\enspace,
%\end{align}
\end{df1}

The perturbation function $\rho$ captures the effect of any small variation of $\mathbf{x}$ with respect to $\mathbf{r}$ on the satisfaction probability of the problem $S_?\utl S_!$ in $\pmc$. For convenience, we call $\mathbf{r}$ the reference of $\rho$.

\subsection{Asymptotic Bounds}

There are various ways to express the asymptotic bounds. We adopt the most basic way: The bounds are given by the so-called \emph{(absolute) condition numbers} \cite{konstantinovetal03}. In Section \ref{sec:relatedwork} we briefly discuss the terminologies of perturbation bounds and condition numbers in the context of related work.

Let $\Delta>0$ represent the perturbation distance of a distribution parameter. In reality, we usually assume $\Delta$ to be a sufficiently small positive number. The following auxiliary definition captures the variation range of $\rho$ with respect to the perturbation distance $\Delta$ of the distribution parameter $\mathbf{x}$.
\begin{df1}\label{df:fluct}
The variation range of $\rho$ with reference $\mathbf{r}$ against $\Delta$ is the set
\begin{align}
    \overline{\rho}(\Delta) =\left\{\rho(\mathbf{v}) ~|~ \|\mathbf{v}-\mathbf{r}\|\leq \Delta, \mathbf{v}\in \mathbf{U} \right\}\enspace.
\end{align}
\end{df1}

It is not hard to see that $\overline{\rho}(\Delta)$ is an interval. The existence of a condition number for $\rho$ depends on the differentiability of $\rho$. The following proposition confirms that $\rho$ enjoys this property in a ``neighborhood'' of $\mathbf{r}$. Recall that we have assumed $|\mathbf{x}|=k$.
\begin{pr1}\label{th:linapprox}
$\rho$ is differentiable at $\mathbf{r}$, namely, $\rho(\mathbf{x})= \mathbf{h}\cdot (\mathbf{x}-\mathbf{r}) + \theta(\mathbf{x}-\mathbf{r})$,
for some $\mathbf{h}\in \mathbb{R}^k$ and $\theta: \mathbb{R}^k\rightarrow \mathbb{R}$ such that $\lim_{\|\mathbf{y}\|\rightarrow 0}\theta(\mathbf{y})/\|\mathbf{y}\|=0$.
\end{pr1}

In other words,
$\mathbf{h}\cdot (\mathbf{x}-\mathbf{r})$ is used as the \emph{linear approximation} of $\rho$ at a point sufficiently close to $\mathbf{r}$,  and we write $\rho(\mathbf{x})\approx\mathbf{h}\cdot  (\mathbf{x}-\mathbf{r})$. Later, we will provide an algorithmic method to determine $\mathbf{h}$. Let $\max(\mathbf{h})=\max\{\mathbf{h}[i] ~|~ 1\leq i\leq |\mathbf{h}|\}$ and $\min(\mathbf{h})=\min\{\mathbf{h}[i] ~|~ 1\leq i\leq |\mathbf{h}|\}$.

\begin{th1}\label{th:condnum}
The \emph{asymptotic bound} of $\rho$ is given by the \emph{condition number}
\begin{align}\label{eq:flucoe}
    \kappa=\lim_{\Delta \rightarrow 0} \sup\left\{\dfrac{x}{\delta} ~|~ x\in \overline{\rho}(\delta),0<\delta\leq \Delta\right\}\enspace.
    %\kappa= & ~ \lim_{\Delta \rightarrow 0} \sup\left\{\dfrac{\rho(\mathbf{v})}{\|\mathbf{v}-\mathbf{r}\|} ~|~ \mathbf{v}\in \mathbf{U},\|\mathbf{v}-\mathbf{r}\|\leq \Delta\right\}\enspace.
\end{align}
Then, the number $\kappa$ exists and, moreover,
\begin{align}\label{eq:fapprx}
    \kappa = \dfrac{1}{2}(\max(\mathbf{h})-\min(\mathbf{h}))\enspace.
\end{align}
%where $\mathbf{h}$ is given in Equation \eqref{eq:linapprox}.
\end{th1}

According to the definition of $\kappa$ in Theorem \ref{th:condnum} (in particular, equation \eqref{eq:flucoe}), mathematically, if the parameter $\mathbf{x}$ in a basic PMC $\pmc$ with reference $\mathbf{r}$ varies an infinitesimally small $\Delta$ from $\mathbf{r}$ in terms of the absolute distance, then the perturbation of the reachability checking result, $\rho(\Delta)$,  is estimated to be within $\pm \kappa\Delta$, where $\kappa$ is the condition number of $\rho$. We test the applicability of such $\kappa$ in experiments in Section 5.

The definition of $\kappa$ captures the sensitivity of $\rho$ to $\mathbf{x}$: How does $\rho$ change if we perturb $\mathbf{x}$? A closely related problem is phrased as this: How much do we have to perturb $\mathbf{x}$ to obtain an approximation of $\rho$---in other words, what is the backward error of $\rho$? The following proposition gives a ``backward'' characterization of the asymptotic bound $\kappa$, which, by its formulation, pursues the infimum of variations of $\mathbf{x}$ (or equivalently, the supremum of their  reciprocals) that can cause the given perturbation of $\rho$.
\begin{pr1}\label{th:condnumback}
$    \kappa = \lim_{x\rightarrow 0} \sup\left\{\delta^{-1}y~|~ 0<y\leq x,\, y\in \overline{\rho}(\delta)\right\}$.
\end{pr1}

In the following, we present a method to compute the linear approximation of $\rho$. We write $\sum_{i=0}^\infty \mathbf{A}^i$ as $\sum\mathbf{A} $. Let
$\mathbf{C}(\mathbf{x})= \mathbf{A}(\mathbf{x})-\mathbf{A}\langle \mathbf{r}\rangle$ and $\mathbf{d}(\mathbf{x})=\mathbf{b}(\mathbf{x})-\mathbf{b}\langle \mathbf{r}\rangle$.
\begin{th1}\label{th:computeh}
Let
\begin{align}
    \mathbf{e}(\mathbf{x})=\sum\mathbf{A}\langle \mathbf{r}\rangle \cdot \mathbf{C}(\mathbf{x})\cdot \sum\mathbf{A}\langle \mathbf{r}\rangle\cdot \mathbf{b}\langle \mathbf{r}\rangle+\sum\mathbf{A}\langle \mathbf{r}\rangle \cdot\mathbf{d}(\mathbf{x})\enspace.
\end{align}
Then, $\rho(\mathbf{x})\approx\iota_{?}\cdot\mathbf{e}(\mathbf{x})$.
\end{th1}

Theorems \ref{th:condnum} and \ref{th:computeh} together provide algorithmic techniques for computing the condition number $\kappa$ for $\pmc$ and the reachability problem $S_?\utl S_!$.

\section{Perturbation Analysis of General PMCs}\label{sec:ana-pmc}

In this section, we generalize the method developed in the previous section from basic PMCs to general PMCs that may have multiple distribution parameters. For general PMCs, perturbations of the parameters may vary either proportionally or independently, yielding two forms of asymptotic bounds, namely, two condition numbers. However, it turns out that the two kinds of bounds coincide.

\subsection{Directional Conditioning}

For general PMCs, we need to handle multiple distribution parameters. The reachability problem $S_?\utl S_!$ in a PMC $\pmc$ is the same as for basic PMCs. In this subsection, we suppose their perturbations are subject to a prescribed ratio, i.e., proportionally. Hence, we associate a function $w: I\rightarrow [0,1]$ to $\mathbf{x}_I$ such that $\sum_{i\in I}w(i)=1$. Such $w$ is called a \emph{direction} of $\mathbf{x}_I$.

To enable the formal treatment, we first define some notations.
For each $i\in I$, let $\mathbf{V}_i=[0,1]^{k_i}$ and  $\mathbf{U}_i = \{\mathbf{v}\in \mathbf{V}_i ~|~ \sum_{j=1}^{k_i }$ $\mathbf{v}[i]=1\}$ where $k_i=|\mathbf{x}_i|$.
If $I=\{i_1,\ldots, i_m\}$, then $\mathbf{V}_I$ denotes the cartesian space $\mathbf{V}_{i_1}\times \ldots \times \mathbf{V}_{i_m}$. Similarly, $\mathbf{U}_I$ is $\mathbf{U}_{i_1}\times \ldots \times \mathbf{U}_{i_m}$. $\mathbf{A}(\mathbf{x}_I)$, $\mathbf{b}(\mathbf{x}_I)$, $\mathbf{A}\langle \mathbf{r}_I \rangle$ and $\mathbf{b}\langle \mathbf{r}_I\rangle$ are natural generalizations of their basic PMC counterparts.
We stress that, unlike $\mathcal{P}(\mathbf{x}_I)$, some variables in $\mathbf{x}_I$ for each $i\in I$ may not appear at $\mathbf{A}(\mathbf{x}_I)$ and $\mathbf{b}(\mathbf{x}_I)$. We can also abbreviate $\mathbf{A}(\mathbf{x}_I)$ and $\mathbf{b}(\mathbf{x}_I)$ as $\mathbf{A}$ and $\mathbf{b}$ if $\mathbf{x}_I$ is clear in the context.

We illustrate these definitions by the example of noisy Zeroconf, whose model is a non-basic PMC. Clearly, the pursuit of the problem ``what is probability of an address collision?'' is equivalent to the problem ``what is probability to avoid an address collision?'' In the second problem, the constraint set is $\{1,\ldots, 5\}$ and the destination set is $\{7\}$. The sequence of parameters is $(x_i, 1-x_i)_{i=1}^4$. Thus,
\begin{align*}
   \mathbf{A}^\mathrm{zf}= \left(
          \begin{array}{ccccc}
            0     &~a~&~0~  &~0~  &~0~  \\
            1-x_1 & 0 & x_1 & 0   & 0   \\
            1-x_2 & 0 & 0   & x_2 & 0   \\
            1-x_3 & 0 & 0   & 0   & x_3 \\
            1-x_4 & 0 & 0   & 0   & 0   \\
          \end{array}
        \right), \quad
    \mathbf{b}^\mathrm{zf}=\left(
          \begin{array}{c}
            1-a\\
            0 \\
            0  \\
            0  \\
            0  \\
          \end{array}
        \right)
\end{align*}

The following definition generalizes Definition \ref{df:rho}.
\begin{df1}
The perturbation function of $\mathbf{x}_I$ for a PMC $\pmc=(\iota, \mathcal{P}(\mathbf{x}_I), \mathbf{r}_I)$ and with respect to the problem $S_?\utl S_!$ such that $S_?, S_!\subseteq S_{\pmc}$ is $\varrho:\mathbf{V}_I\rightarrow [-1,1]$ such that
\begin{align}\label{eq:varrho}
    \varrho(\mathbf{x}_I)
    %= &~\sum_{i=1}^{n_?} \iota[i]\sum_{j=0}^\infty (\mathbf{A}_{\mathbf{x}^\ast_{I,?}}^j\cdot \mathbf{b}_{ \widehat{\mathbf{x}}^\ast_{I,!}}-\mathbf{A}^j\cdot \mathbf{b})[i] \nonumber\\
    = & ~ \iota_?\cdot \sum_{j=0}^\infty (\mathbf{A}(\mathbf{x}_I)^j\cdot \mathbf{b}(\mathbf{x}_I)-\mathbf{A}\langle \mathbf{r}_I\rangle^j\cdot \mathbf{b}\langle \mathbf{r}_I\rangle)\enspace.
\end{align}
%where $g(\mathbf{x}_I)$ is given by
%\begin{align}\label{eq:g}
    %g(\mathbf{x}_I)=& ~ \sum_{j=0}^\infty \mathbf{A}_{\mathbf{x}^\ast_{I,?}}^j\cdot \mathbf{b}_{ \widehat{\mathbf{x}}^\ast_{I,!}}\enspace.
%\end{align}
\end{df1}

The perturbation function $\varrho$ captures the effect of the small variation of $\mathbf{x}_i$ with respect to $\mathbf{r}_i$ for each $i\in I$ on the reachability problem $S_?\utl S_!$ in $\pmc$. We call vectors in $\mathbf{r}_I$ references of $\varrho$.
The definition below generalizes Definition \ref{df:fluct}.
\begin{df1}
The $w$-direction variation range of $\varrho$ with reference in $\mathbf{r}$ against $\Delta$ is the set
\begin{align}
    \overline{\varrho}_w(\Delta) = \{\varrho(\mathbf{v}_I) ~|~ \| \mathbf{v}_i-\mathbf{r}_i\|\leq w(i)\Delta,\, \mathbf{v}_i\in \mathbf{U}_i,\, i\in I\}\enspace,
\end{align}
where $\mathbf{v}_I=(\mathbf{v}_i)_{i\in I}$.
\end{df1}

Let $\mathbf{x}_I-\mathbf{r}_I$ be the sequence $(\mathbf{x}_i-\mathbf{r}_i)_{i\in I}$, supposing $|\mathbf{x}_i|=|\mathbf{r}_i|$ for each $i\in I$.
Let $\|\mathbf{x}_I\|$ be $\sum_{i\in I}\|\mathbf{x}_i\|$. Similar to $\rho$, the following proposition holds for $\varrho$.
\begin{pr1}\label{th:linapproxg}
$\varrho$ is differentiable at $\mathbf{r}_I$, namely,
$\varrho(\mathbf{x}_I)= \sum_{i\in I} \mathbf{h}_i\cdot (\mathbf{x}_i-\mathbf{r}_i) + \theta'(\mathbf{x}_I-\mathbf{r}_I)$,
for some $\mathbf{h}_i\in \mathbb{R}^k$ ($i\in I$) and $\theta': \mathbb{R}^{k|I|}\rightarrow \mathbb{R}$ such that $\lim_{\|\mathbf{y}_I\|\rightarrow 0}\theta'(\mathbf{y}_I)/\|\mathbf{y}_I\|=0$.
\end{pr1}

We write $\varrho(\mathbf{x}_I)\approx \sum_{i\in I} \mathbf{h}_i\cdot (\mathbf{x}_i-\mathbf{r}_i)$ and call $\sum_{i\in I} \mathbf{h}_i\cdot (\mathbf{x}_i-\mathbf{r}_i)$ the \emph{linear approximation} of $\varrho$ at $\mathbf{r}_I$. The following theorem generalizes Theorem \ref{th:condnum}.
\begin{th1}\label{th:condnumdi}
The \emph{$w$-direction asymptotic bound} of $\varrho$ is given by the \emph{directional condition number}
\begin{align}\label{eq:flucoeg}
    \kappa_w=\lim_{\Delta \rightarrow 0} \sup\left\{\dfrac{x}{\delta} ~|~ x\in \overline{\varrho}_w(\delta), 0<\delta\leq \Delta\right\}\enspace.
\end{align}
Then, $\kappa_w$ exists and, moreover,
\begin{align}\label{eq:fapprxg}
    \kappa_w = \dfrac{1}{2}\sum_{i\in I}w(i)(\max(\mathbf{h}_i)-\min(\mathbf{h}_i))\enspace.
\end{align}
%where each $\mathbf{h}_i$ is given in Equation \eqref{eq:linapproxg}.
\end{th1}

If the distribution parameters vary a small $\Delta$ in the direction $w$, then we can estimate the perturbation of $\rho$ as $\pm k_w\Delta$. In the case that $w(i)=1/|I|$ for each $i\in I$, such $k_w$ is called a \emph{uniform} condition number.
Like the asymptotic bounds for basic PMCs, a ``backward'' characterization of $\kappa_w$ also exists, as follows.
\begin{pr1}\label{th:condnumbackdi}
    $\kappa_w = \lim_{x\rightarrow 0} \sup \left\{\delta^{-1}y~|~ 0<y\leq x,\, y \in \overline{\varrho}_w(\delta)\right\}.$
\end{pr1}

We provide a method to compute the linear approximation of $\varrho$.
We define two specific parametric matrices: $\mathbf{C}(\mathbf{x}_I)= \mathbf{A}(\mathbf{x}_I)-\mathbf{A}\langle \mathbf{r}_I\rangle$ and $\mathbf{d}(\mathbf{x}_I)=\mathbf{d}(\mathbf{x}_I)-\mathbf{d}\langle \mathbf{r}_I\rangle$.
 We have the following generalized theorem of Theorem \ref{th:computeh}.
\begin{th1}\label{th:computehg}
For each $i\in I$, let
\begin{align}
    \mathbf{e}(\mathbf{x}_I)=\sum\mathbf{A}\langle \mathbf{r}_I\rangle \cdot \mathbf{C}(\mathbf{x}_I)\cdot \sum\mathbf{A}\langle \mathbf{r}_I\rangle\cdot\mathbf{b}\langle \mathbf{r}_I\rangle+ \sum\mathbf{A}\langle \mathbf{r}_I\rangle \cdot\mathbf{d}( \mathbf{x}_I)\enspace.
\end{align}
Then, $\varrho(\mathbf{x}_I)\approx \iota_{?}\cdot\mathbf{e}( \mathbf{x}_I)$.
\end{th1}

Theorems \ref{th:condnumdi} and \ref{th:computehg} together provide an algorithmic method for computing the directional condition number $\kappa_w$ for $\pmc$ and the reachability problem $S_?\utl S_!$.

\subsection{Parameter-wise Conditioning}

The parameter-wise perturbation analysis handles the independent variations of distribution parameters. In this case, to facilitate perturbation estimation, we expect to obtain a condition number for each distribution parameter. It turns out that the parameter-wise analysis can be reduced to the directional analysis.

We use $\mathbf{r}_I[i:=\mathbf{v}]$ denote the sequence of vectors obtained by replacing the $i$th vector in $\mathbf{r}_I$ by $\mathbf{v}$.
\begin{df1}
The variation range of $\varrho$ \emph{projected} at $\mathbf{x}_i$ against $\Delta$ is the set
\begin{align}
    \overline{\varrho}_i(\Delta) = \{\varrho(\mathbf{r}_I[i:=\mathbf{v}]) ~|~ \| \mathbf{v}-\mathbf{r}_i\|\leq \Delta,\, \mathbf{v}\in \mathbf{U}_i\}\enspace.
\end{align}
The \emph{asymptotic bound} of $\varrho$ \emph{projected} at $\mathbf{x}_i$ is given by the condition number
\begin{align}\label{eq:flucoeg}
    \kappa_i=\lim_{\Delta \rightarrow 0} \sup\left\{\dfrac{x}{\delta} ~|~ x\in \overline{\varrho}_i(\delta), 0<\delta\leq \Delta\right\}\enspace.
\end{align}
\end{df1}

Let $w_i$ be the direction such that $w_i(i)=i$ and $w_i(j)=0$ for each $j\in I\backslash \{i\}$. It is easy to see that $\overline{\varrho}_i$ (resp.~$\kappa_i$) is just $\overline{\varrho}_{w_i}$ (resp.~$\kappa_{w_i}$). It means that parameter-wise bounds are special cases of directional bounds. Moreover, the following theorem states that any set of parameter-wise condition numbers conforms to a specific directional condition number.
\begin{th1}\label{th:link}
Let $\Delta=\sum_{i\in I} \Delta_i$ and $w(i)= \Delta_i/ \Delta$ for each $i\in I$. Then,
$ \sum_{i\in I} \kappa_i\Delta_i = \kappa_w \Delta$.
\end{th1}

If the direction of perturbation may not be known in advance, it is more useful to present the set of parameter-wise bounds. Theorem \ref{th:link} provides a mathema\-tical characterization for parameter-wise perturbations in terms of directional perturbations.

\section{Experiments}\label{sec:experiment}

We evaluate by experiments, how well the condition numbers capture possible perturbations of reachability probabilities for some PMCs under consideration. Recall that the outcome of the reachability checking algorithm for a PMC consists of two parts, namely, a referential probabilistic result and one or more condition numbers (see Table \ref{tb:mcproblem}). The probabilistic result is computed by a conventional numerical model checking algorithm.
%In particular, the probability result is given by the equation
%$    p= \iota_{?}\cdot \mathbf{p}+\iota[n_!]+\ldots+\iota[|\iota|] $
%where $\mathbf{p}$ is given in equation \eqref{eq:preachability}.
For the  problems considered in this section, only a single condition number will be returned. The condition number is calculated by the method presented in the previous sections.

Our experiments proceed as follows. (i) We specify a reachability problem for a PMC $\pmc$ and compute the referential probabilistic result $p$ and \emph{one} condition number $\kappa$ for the problem, although multiple condition numbers may be required in other contexts. (ii) By deliberately assigning concrete probability distributions to the distribution parameter(s) of $\pmc$, we construct several potential non-parametric models $\mc_j$ with sufficiently small statistical distances from $\pmc$. (iii) We compute an actual probabilistic result $p_j$ for each $\mc_j$ and calculate the actual distance $\Delta_j$ between the reference(s) in $\pmc$ and the corresponding distribution(s) in $\mc_j$. (iv) We compare $p-p_j$, the difference between the referential result and an actual result, and $\pm \kappa\Delta_j$, the perturbation estimation.

\begin{table}[t]
\centering
\caption{Experimental data of noisy Zeroconf ($\times 10^{-3}$)}\label{tb:expdatazf}
\begin{tabular}{cccccc}
 \hline
  \multirow{2}{*}{~Model~}& \multirow{2}{*}{$x_i$} & \multirow{2}{*}{~Probability~}  & \multirow{2}{*}{~Distance~} & ~Condition~&~Variation \vspace{-.3em} \\
  &&&& Number & Range \\
  \hline \noalign{\smallskip}
  ~$\pmc^{\mathrm{zf}}$~ & ~$750$~ & $999.024$ &  - & $7.797$ & - \\
  ~$\mc^{\mathrm{zf}}_1$~ & $749$ & $-.016$  & $2$ & - & $\pm .016$ \\
  $\mc^{\mathrm{zf}}_2$ & $752$ & $+ .031$  & $4$ & - & $\pm .031$ \\
  $\mc^{\mathrm{zf}}_3$ & $747$ & $ -.048$  & $6$  & - & $\pm .047$ \\
  \noalign{\smallskip} \hline
\end{tabular}
\end{table}

We performed experiments for the examples of noisy Zeroconf and hopping frog (the PMCs $\pmc^\mathrm{zf}$ and $\pmc^\mathrm{fg}$) in Matlab$^\circledR$ \cite{matlab12}. Although Matlab is not a specialized tool for probabilistic model checking, it provides convenient numerical and symbolic mathematical operations that are necessary to compute the perturbation function. For the easier calculations in the noisy Zeroconf example, we let $a$ in $\mathcal{P}_\ast^\mathrm{zf}$, the parametric transition matrix of $\pmc^\mathrm{zf}$, be $0.2$. Moreover, for simplicity, it is assumed that each probe message and its reply are affected by the same channel noise level, namely, that the four distribution parameters of $\pmc^\mathrm{zf}$ are perturbed in a uniform direction. Several MCs in both experiments are generated by assigning different distributions to the distribution parameters in their PMCs.
Because the infinite matrix series $\sum_{i=0}^\infty \mathbf{A}^j$ cannot be computed directly, we adopt an approximation by taking the sum of the first hundred items in each series encountered. There was no significant truncation error involved the numerical calculations in our experiments. We test the reachability problems $\{1,\ldots,5\}\utl \{7\}$ in the first experiment (which states ``what is the probability to avoid an IP collision?'') and $\{1,2\}\utl\{4\}$ in the second one (which states ``what is the probability for the frog to reach the fourth rock without landing on the third rock?'').
The experimental data are summarized in Tables \ref{tb:expdatazf} and \ref{tb:expdatafg}, respectively. In Table \ref{tb:expdatazf}, the distance of the perturbed models to the PMC increases. We observe that the condition number accounts for the result nicely if the perturbed distance is smaller than $0.006$. When the distance exceeds $0.006$, the perturbation of the probabilistic result may exceed the variation range. In Table \ref{tb:expdatafg}, several perturbed models with the distance $0.004$ to the PMC are presented. The data also demonstrate that the condition number bounds the reachability perturbation between a perturbed model and the PMC to a satisfactory degree. In particular, the difference between the result for $\mc^\mathrm{fg}_4$ (resp., $\mc^\mathrm{fg}_5$) and the referential result overlaps with the positive (resp., negative) predicted bound.

\begin{table}[t]
\centering
\caption{Experimental data of hopping frog ($\times 10^{-3}$)}\label{tb:expdatafg}
\begin{tabular}{c ccccc}
  \hline
   \multirow{2}{*}{~Model~} & \multirow{2}{*}{Distribution} & \multirow{2}{*}{~Probability~}  & \multirow{2}{*}{~Distance~} & ~Condition~  & ~Variation~ \vspace{-.3em} \\
  &&&& Number & Range \\
  \hline \noalign{\smallskip}
  ~$\pmc^{\mathrm{fg}}$~  & ~$(375,125,250, 250)~$ & $500.000$ &  - & $312.500$ & - \\
  ~$\mc^{\mathrm{fg}}_1$~  & $(374,124,251,251)$ & $0$  & $4$  & - & $\pm 1.250$ \\
  $\mc^{\mathrm{fg}}_2$  & $(374,124,250,252)$ & $+.623$  & $4$ & - & $\pm 1.250$ \\
  $\mc^{\mathrm{fg}}_3$  & $(377,125,248,250)$ & $ +.627$ & $4$ & - & $\pm 1.250$ \\
  $\mc^{\mathrm{fg}}_4$  & $(377,125,250,248)$ & $-.627 $ & $4$ & - & $\pm 1.250$ \\
  %$\mc^{\mathrm{fg}}_4$  & $(377,125,249,249)$ & $0 $ & $4$ & - & $\pm 1.250$ \\
  $\mc^{\mathrm{fg}}_4$  & $(375,125,248,252)$ & $+1.250 $ & $4$ & - & $\pm 1.250$ \\
  $\mc^{\mathrm{fg}}_5$  & $(375,125,252,248)$ & $-1.250 $ & $4$ & - & $\pm 1.250$ \\
  \noalign{\smallskip} \hline
\end{tabular}
\end{table}

In short, we observe from the experiments that condition numbers adequately, although not rigorously, predict the bounds of the reachability checking results for probabilistic models under small perturbations.

\section{Discussion and Related Work}\label{sec:relatedwork}

The pursuit of perturbation bounds for MCs can be traced back to the 1960's. Schweitzer \cite{schweitzer68} gave the first perturbation bound, namely, an absolute condition number for the stationary distribution of an MC against its fundamental matrix (which is defined by the transition matrix of the MC), and this motivated a variety of subsequent work. Cho and Meyer \cite{chomeyer00} provided an excellent overview for various bounds of stationary distributions (all of which are condition numbers) up to the time of their publication, whilst more recent papers \cite{solanvieille03,heigergott08} shed light on new definitions and techniques for perturbation bounds. In spite of its relatively long history, to the best of our knowledge, the present paper is the first paper that studies the perturbation problem in quantitative verification. Moreover, our approach is different from most of the works on the perturbation analysis for MCs in that, instead of formulating the bounds in terms of mathematically meaningful components, we adopt numerical computation to approximate the bounds. Therefore, our work is in mid of a broader branch of perturbation theory for numerical linear algebra \cite{trefethenbau97}, the goal of which is to investigate the sensitivity of a matrix-formulated problem with respect to one or more perturbed components in its formulation, and to provide various forms of perturbation bounds for the solution to the problem. One important group of such bounds is called asymptotic bounds (also called linear local bounds), which is further divided into two subgroups, namely, absolute condition numbers and relative condition numbers. Both subgroups of condition numbers have their own significance---absolute condition numbers enjoy a more elegant mathematical formulation and are easier to employ for practical problems, whilst relative ones are more important to the floating point arithmetic implemented in every computer, which is affected by relative rather than absolute errors. A detailed classification of these bounds is found in Konstantinov \emph{et al.} \cite{konstantinovetal03} (Chapters 1 and 2). The condition numbers that we pursue in the present paper are absolute ones and we leave the analysis of our problem based on other kinds of bounds to future work.

Quantitative verification of Markov models with various formulations of uncertainty is a recently active field of research. Daws \cite{daws04} proposed a symbolic PCTL model checking approach in which concrete or abstract transition probabilities in his parametric variant of a discrete-time MC are viewed as letters in an alphabet of a finite automaton. As such, the probability measure of a set of paths satisfying a formula is computed symbolically as a regular expression on that alphabet, which is further evaluated to its exact rational value when transition probabilities are rational symbolic expressions of variables. Hahn \emph{et al.} \cite{hahnetal11} improved the approach of Daws for reachability checking (i.e., PCTL formulae without nested probability operators) by carefully intertwining the computation procedure and evaluation procedure of Daws. By definition, their parametric variants of MCs are more general than ours because they allow abstract transition probabilities to be expressed by rational symbolic expressions. But in order to introduce a metric to measure the perturbations for our PMCs, we let abstract transition probabilities be expressed as single variables. Another and more important difference is that, instead of pushing the symbolic computation to an extreme as they did, we calculate numbers in symbolic expressions numerically as in ordinary mathematical calculations.

Another group of research works addresses the undetermined transition probabilities in MCs by specifying their interval values. Sen \emph{et al.} \cite{senetal06} considered two semantic interpretations for such models, which are either classes of \mbox{MC} or generalizations of Markov Decision Processes (MDPs). In the first interpretation, the PCTL model checking problem is to search for an MC within the MC class such that a PCTL formula is satisfied; in the second one, the problem can be reduced to a corresponding MDP of exponential size. \mbox{Benedikt \emph{et al.}} \cite{benediktetal} consi\-dered the LTL model checking problem for the same models, which they defined as the search of an MC that meets the model constraint and optimizes the probability of satisfying an LTL formula. However, in our perturbation approach we specify a metric to measure the perturbed distances of the models but not their perturbed boundaries in terms of interval transition probabilities.

There have also been attempts to study perturbation errors in realtime systems, in particular, timed automata. For example, Alur \emph{et al.} \cite{aluretal05} defined a perturbed semantics for timed automata whose clocks might skew at some very small rates. They showed that if an automaton has a single clock, then the language accepted by it under the perturbed semantics is accepted by an equivalent deterministic automaton under the standard semantics. Bouyer \emph{et al.} \cite{bouyeretal06} provided another time perturbation notion, which expresses not the perturbations of the clock rates but those of the clock constraints. They developed model checking techniques for $\omega$-regular properties based on their novel semantic relation, which captures---as argued---the intuition ``whether the considered property holds for the same model implemented in a sufficient (but not infinitely) fast hardware''.

%Still another different problem, which studies realtime systems with general dynamic nature, was addressed by Chen \emph{et al} \cite{chenetal09}, who proposed an LTL model checking method for a continuous-time inhomogeneous (or non-stationary) MC, the inhomogeneity of which is predefined by an integrable function.

%Sensitivity of computable problems to their input data are also explored in other contexts. Misailovic \emph{et al} \cite{misailovicetal11} introduced a framework of approximate correctness for programs, one utility of which is to analyze the robustness of programs with disturbed input, namely, whether small perturbations to the input lead to small variations of the output only. Heimdahl \emph{et al } \cite{heimdahletal02} employed model checking techniques in software deviation analysis of control systems.

\section{Conclusions}\label{sec:conclu}

Motivated by the pervasive phenomena of perturbations in the modeling and verification of real-life probabilistic systems, we studied the sensitivity of constrained reachability probabilities of those systems---which are modeled by parametric variants of discrete-time MCs---to perturbations of their distribution \mbox{parameters}. Our contribution is a method to compute the asymptotic bounds in terms of absolute condition numbers for characterizing the sensitivity. We also conducted experiments to demonstrate the practical adequacy of the computation method.

This paper is an initial step towards investigating the sensitivity and bounds for quantitative verification of perturbed systems, and we may identify several interesting directions for further research. First, reachability, in spite of its fundamental status in model checking, captures only a narrow group of practical verification problems (particularly in the probabilistic domain) and, therefore, it is desirable to extend the present method to accommodate the general model checking problems formalized, for instance, in LTL formulas. Second, we adopt the norm of absolute distance to measure the distance between two probabi\-lity distributions; however, there exist other distance measures that are useful for problems in some specific domains. For example, the well-known Kullback-Leibler divergence is widely adopted in information theory \cite{coverthomas91}. Finally, condition numbers are among several other forms of perturbation bounds. An in-depth comparison of their pros and cons is left to future work.

\nocite{baierkatoen08}

\bibliographystyle{splncs}
\bibliography{Bibil}

\section*{Appendix: Proof Details}

\begin{proof}[Proof of Lemma \ref{le:preachability}]
See Theorem 10.15 and its subsequent remark at pp.762-764 in \cite{baierkatoen08}.
\end{proof}

\begin{proof}[Proof of Proposition \ref{th:linapprox}]
We refer this proof to the constructive (and independent) proof of Theorem \ref{th:computeh} below.
\end{proof}

\begin{proof}[Proof of Theorem \ref{th:condnum}]
Let
\begin{align*}
    \kappa'=  \dfrac{1}{2}(\max(\mathbf{h})-\min(\mathbf{h}))
    =  \dfrac{1}{2}(\mathbf{h}[i_1]-\mathbf{h}[i_2])
\end{align*}
for some $i_1,i_2$, and our goal is to show that $\kappa$ exists and $\kappa=\kappa'$. Let $|\mathbf{h}|=k$.

First we give a claim. Let $\mathbf{q}\in [-1,1]^{k}$ such that $\sum_{i=1}^k \mathbf{q}=0$ and $\|\mathbf{q}\|=1$. We claim that
\begin{align*}
    \mathbf{h}\cdot\mathbf{q}\leq \kappa'\enspace.
\end{align*}
The proof of the claim is straightforward.

Let $\varepsilon>0$. According to Proposition \ref{th:linapprox}, choose $\Delta>0$ such that if $0<\|\mathbf{x}\|=\delta \leq \Delta$ then
\begin{align*}
    \left|\mathbf{h}\cdot\mathbf{x}-\rho(\mathbf{y} +\mathbf{r})\right|\leq  \dfrac{\delta\varepsilon}{2}\enspace.
\end{align*}
Let $\mathbf{y}=\delta\mathbf{q}$, then
\begin{align*}
    \left|\mathbf{h}\cdot\mathbf{q} -
    \dfrac{\rho(\mathbf{y}+\mathbf{r})}
    {\|\mathbf{y}\|}\right| = \left| \dfrac{\mathbf{h}\cdot\mathbf{x}-\rho(\mathbf{y} +\mathbf{r})}{\|\mathbf{y}\|} \right|\leq \dfrac{\varepsilon}{2} \enspace;
\end{align*}
in particular,
\begin{align*}
 \left|\kappa'-\dfrac{\rho(\delta\mathbf{e}_{i_1,i_2}+\mathbf{r})}
    {\|\delta\mathbf{e}_{i_1,i_2}\|}\right|\leq \dfrac{\varepsilon}{2}\enspace,
\end{align*}
where $\mathbf{e}_{i,j}$ is the vector such that the $i$th (resp.~$j$th) entry in $\mathbf{e}_{i,j}$ is $1/2$ (resp.~$-1/2$) and the other entries (if any) are all zero. Since $\Delta$ is supposed to be sufficiently small, $\delta\mathbf{e}_{i_1,i_2}+\mathbf{r}\in \mathbf{U}$. Then,
$$\dfrac{\rho(\delta\mathbf{e}_{i_1,i_2}+\mathbf{r})}{\delta}\in \left\{\dfrac{x}{\delta} ~|~ x\in \overline{\rho}(\delta), 0<\delta\leq\Delta\right\} \enspace.$$
Thus,
\begin{align*}
    \kappa'-\varepsilon\leq \sup\left\{\dfrac{x}{\delta} ~|~ x\in \overline{\rho}(\delta), 0<\delta\leq\Delta\right\} \enspace.
\end{align*}

On the other hand, given $0<\delta'<\Delta$, choose $\mathbf{y}'$ such that $\|\mathbf{y}'\|=\delta'$, $\mathbf{y}'+\mathbf{r}\in \mathbf{U}$, and $\sup(\overline{\rho}(\delta'))\leq \rho(\mathbf{y}'+\mathbf{r})+\varepsilon\delta'/2$. Without loss of generality, we suppose $\mathbf{y}' =\delta'\mathbf{q}'$. Thus,
\begin{align*}
     \mathbf{h}\cdot\mathbf{q}'+\varepsilon \geq \dfrac{\rho(\mathbf{y}'+\mathbf{r})}{\delta'}+\dfrac{\varepsilon}{2} \enspace.
\end{align*}
Thus, $\kappa'+\varepsilon \geq \dfrac{\sup(\overline{\rho}(\delta))}{\delta}$ for any $\delta$ such that $0<\delta\leq\Delta$. Hence,
\begin{align*}
    \kappa'+\varepsilon\geq \sup\left\{\dfrac{x}{\delta} ~|~ x\in \overline{\rho}(\delta), 0<\delta\leq\Delta\right\} \enspace.
\end{align*}

Therefore, we have proved that $\kappa$ exists and $\kappa=\kappa'$.
\end{proof}

\begin{proof}[Proof of Proposition \ref{th:condnumback}]
We suffice to show the following two propositions:
\begin{enumerate}
  \item for each $\varepsilon>0$ there exists $x>0$ such that $y/\delta< \kappa+ \varepsilon$ whenever $0<y\leq x$ and $y\in \overline{\rho}(\delta)$;
  \item for each $\varepsilon>0$ and $x>0$ there exists $0<y\leq x$ and $\delta>0$ such that $y\in \overline{\rho}(\delta)$ and $\kappa<y/\delta+ \varepsilon$.
\end{enumerate}
On the other hand, by the definition of $\kappa$,
\begin{enumerate}
  \item for each $\varepsilon>0$ there exists $\Delta>0$ such that $y/\delta< \kappa+ \varepsilon$ whenever $0<\delta\leq \Delta$ and $y\in \overline{\rho}(\delta)$;
  \item for each $\varepsilon>0$ and $\Delta>0$ there exists $0<\delta\leq \Delta$ and $y>0$ such that $y\in \overline{\rho}(\delta)$ and $\kappa<y/\delta+ \varepsilon$.
\end{enumerate}
It holds that given any $\Delta>0$ we can choose $x$ such that if $0<y\leq x$ and $y\in \overline{\rho}(\delta)$ then $0 <\delta\leq\Delta$; conversely, given any $x>$ we can choose $\Delta$ such that if $0<\delta\leq \Delta$ and $y\in \overline{\rho}(\delta)$ then $0 <y\leq x$. Therefore, it can be verified that the two sets of propositions are equivalent.
\end{proof}

\begin{proof}[Proof of Theorem \ref{th:computeh}]$\rho(\mathbf{x})$ actually defines a system of non-linear (i.e{.} multi-variable polynomial) series.
Given a multi-variable polynomial series, the order of a term in the series is the summation of the exponents of all variants in it. The smallest order of all terms is called least order of the series. We write $\mathbf{A}\langle \mathbf{r}\rangle$ and $\mathbf{b}\langle \mathbf{r}\rangle$ as $\mathbf{A}$ and $\mathbf{b}$, respectively.
By the definition of $\mathbf{C}(\mathbf{x})$ and $\mathbf{d}(\mathbf{x})$, we rewrite $\rho$ as follows:
\begin{align*}
    & \rho(\mathbf{x}+\mathbf{r})\\
    = ~&  \iota_{?} \cdot \left(\sum_{i=0}^\infty\left(\mathbf{A} + \mathbf{C}(\mathbf{x})\right)^i \cdot \left( \mathbf{b}+\mathbf{d}(\mathbf{x})\right)- \sum_{i=0}^\infty\mathbf{A}\cdot \mathbf{b}\right) \\
    %= ~& \iota_{?} \cdot \left(\sum_{i=0}^\infty  \mathbf{A}^i \cdot \mathbf{d}_{\widehat{\mathbf{y}}^\ast _{!}} +\sum_{j=0}^\infty \sum_{j=0}^{i}  \mathbf{A}^{i-j}\cdot \mathbf{D}_{\mathbf{y}^\ast_{?}}\cdot \mathbf{A}^{j}\cdot \mathbf{b}\right)+\varphi\\
    =~& \iota_{?} \cdot \left(\underbrace{\sum_{i=0}^\infty  \mathbf{A}^i \cdot \mathbf{d}(\mathbf{x})}_{\psi_1} + \underbrace{\sum_{i=0}^\infty  \mathbf{A}^{i}\cdot\mathbf{C}(\mathbf{x})\cdot\sum_{i=0}^\infty  \mathbf{A}^{i}\cdot \mathbf{b}}_{\psi_2}\right)+\varphi \enspace,
\end{align*}
for some $\varphi$. We see that $\varphi$ is either $0$ or a of polynomial whose least order is not smaller than $2$. Thus,
\begin{align*}
    \lim_{\|\mathbf{x}\|\rightarrow 0}\dfrac{\varphi}{\|\mathbf{x}\|}=0 \enspace.
\end{align*}
Let $\mathbf{h}\cdot\mathbf{x}=\iota_{?}\cdot(\psi_1+\psi_2)$, and we obtain the equation in Proposition \ref{th:linapprox}, namely
\begin{align*}
    \rho(\mathbf{x}+\mathbf{r})\approx \iota_{?}\cdot (\psi_1+\psi_2) \enspace.
\end{align*}
\end{proof}

\begin{proof}[Proof of Proposition \ref{th:linapproxg}]
We refer this proof to the proof of Theorem \ref{th:computehg} below.
\end{proof}

\begin{proof}[Proof of Theorem \ref{th:condnumdi}]
The proof is a generalization of the proof of Theorem \ref{th:condnum}. For any direction $w$, let
\begin{align*}
    \kappa'_w = \dfrac{1}{2}\sum_{i\in I}w(i)(\max(\mathbf{h}_i)-\min(\mathbf{h}_i))=\dfrac{1}{2}\sum_{i\in I}w(i)(\mathbf{h}_i[j_i^1]-\mathbf{h}_i[j_i^2])\enspace,
\end{align*}
for some $j_i^1,j_i^2$ for each $i\in I$. Our goal is to show that $\kappa_w$ exists and $\kappa_w=\kappa'_w$. Note that $|\mathbf{h}_i|=k$ for each $i$.

We first give a claim. Let $\mathbf{q}_i\in [-1,1]^{k}$ such that, for each $i\in I$, $\sum_{j=1}^k \mathbf{q}_i[j]=0$ and $\|\mathbf{q}_i\|=1$. We claim that
\begin{align*}
   \sum_{i\in I} w(i)(\mathbf{h}_i\cdot\mathbf{q}_i)\leq \kappa'|I|\enspace.
\end{align*}
The claim is not hard to be verified.

Let $\varepsilon>0$. According to Proposition \ref{th:linapproxg}, choose $\Delta>0$ such that if $0<\|\mathbf{x}_{i}\| =\delta w(i) \leq \Delta w(i)$ for any $i\in I$, then
\begin{align*}
    \left|\sum_{i\in I}\mathbf{h}_i\cdot\mathbf{x}_{i}-\varrho(\mathbf{x}_{I} +\mathbf{r}_I)\right|\leq  \dfrac{\delta\varepsilon}{2} \enspace.
\end{align*}
where $\mathbf{x})_I$ is $(\mathbf{x}_i)_{i\in I}$. Let $\mathbf{x}_{i}=\delta w(i)\mathbf{q}_i$ for each $i\in I$. Thus %(since $\|\mathbf{x}_{\Delta, I}\|=\sum_{i\in I}\|\mathbf{x}_{\Delta, i}\|=2 r\|\mathbf{w}\|$)
\begin{align*}
    \left| {\sum_{i\in I} w(i)(\mathbf{h}_i\cdot\mathbf{q}_i)}-
    \dfrac{\varrho(\mathbf{x}_{ I}+\mathbf{r}_I)}
    {\delta}\right| = \left|\dfrac{\sum_{i\in I} \mathbf{h}_i\cdot\mathbf{x}_{i}-\varrho(\mathbf{x}_{I} +\mathbf{r}_I)}{\delta} \right|\leq \dfrac{\varepsilon}{2} \enspace;
\end{align*}
in particular, let $\mathbf{e}_I=(\mathbf{e}_{j_i^1,j_i^2})_{i\in I}$ where $\mathbf{e}_{j_1^1,j_i^2}$ is the vector such that the $j_i^1$th (resp.~$j_i^2$th) entry in it is $1/2$ (resp.~$-1/2$) and the other entries (if any) are all zero (thus, $\mathbf{e}_{j_i^1,j_i^2}$ is a particular instance of $q_i$); then
\begin{align*}
 \left|\kappa'_w-\dfrac{\varrho(\delta w\cdot\mathbf{e}_{I}+\mathbf{r}_I)}
    {\delta}\right|\leq \dfrac{\varepsilon}{2}\enspace.
\end{align*}
Since $\delta$ is sufficiently small, $\delta w\cdot\mathbf{e}_{I}+\mathbf{r}_I\in \mathbf{U}_I$. Then
$$\dfrac{\varrho(\delta w\cdot \mathbf{e}_{I}+\mathbf{r}_I)}{\delta}\in \left\{\dfrac{x}{\delta} ~|~ x\in \overline{\varrho}_w(\delta), 0<\delta\leq\Delta\right\} \enspace.$$
Thus,
\begin{align*}
    \kappa'_w-\varepsilon\leq \sup\left\{\dfrac{x}{\delta} ~|~ x\in \overline{\varrho}_w(\delta), 0<\delta\leq\Delta\right\} \enspace.
\end{align*}

On the other hand, given $0<\delta'\leq \Delta$, choose $\mathbf{x}_{I}'=(\mathbf{x}_{i}')_{i\in I}$ such that $\|\mathbf{x}_{i}'\| =\delta' w(i)$ for each $i\in I$, $\mathbf{x}_{I}'+\mathbf{r}_I\in \mathbf{U}_I$, and $\sup(\overline{\varrho}(\delta'))\leq\rho(\mathbf{x}_{I}'+\mathbf{r}_I)+\varepsilon\delta'/2$. Without loss of generality, we suppose $\mathbf{x}_{i}' =\delta'\mathbf{q}'_i$ for each $i\in I$. Thus,
\begin{align*}
     \sum_{i\in I}\mathbf{h}_i\cdot\mathbf{q}'_i+\varepsilon \geq \dfrac{\varrho(\mathbf{x}_{I}'+\mathbf{r}_I)}{\delta'}+\dfrac{\varepsilon}{2} \enspace.
\end{align*}
Thus, $\kappa'_w+\varepsilon \geq \dfrac{\sup(\overline{\varrho}_w(\delta'))}{\delta'}$. Hence,
\begin{align*}
    \kappa'_w+\varepsilon\geq \sup\left\{\dfrac{x}{\delta} ~|~ x\in \overline{\varrho}_w(\delta),\, 0<\delta\leq\Delta\right\} \enspace.
\end{align*}

Therefore, we have proved that $\kappa_w$ exists and $\kappa_w=\kappa'_w$.
\end{proof}

\begin{proof}[Proof of Proposition \ref{th:condnumbackdi}]
The proof is an immediate generalization of the proof of Proposition \ref{th:condnumback}.
\end{proof}

\begin{proof}[Proof of Theorem \ref{th:computehg}]
The proof of this theorem is a generalization of that of Theorem \ref{th:computeh}.
 We write $\mathbf{A}\langle \mathbf{r}_I\rangle$ and $\mathbf{b}\langle \mathbf{r}_I\rangle$ as $\mathbf{A}$ and $\mathbf{b}$, respectively.
By the definition of $\mathbf{C}(\mathbf{x}_I)$ and $\mathbf{d}(\mathbf{x}_I)$, we rewrite $\varrho$ as follows:
\begin{align*}
    & \varrho(\mathbf{x}_I+\mathbf{r}_I)\\
    %= ~ & \iota_{?} \cdot (g(\mathbf{x}_I+\mathbf{r})-g(\mathbf{r}_I)) \\
    = ~&  \iota_{?} \cdot \sum_{i\in I}\sum_{j=0}^\infty\left((\mathbf{A}+ \mathbf{C}(\mathbf{x}_I))^j \cdot ( \mathbf{b}+\mathbf{d}(\mathbf{x}_I))- \mathbf{A}^j\cdot \mathbf{b}\right)\\
    %= ~& \iota_{?} \cdot \left(\sum_{i=0}^\infty  \mathbf{A}^i \cdot \mathbf{0}_{\widehat{\mathbf{x}}^\ast _{!}}^{(n_?)} +\sum_{j=0}^\infty \sum_{j=0}^{i}  \mathbf{A}^{i-j}\cdot \mathbf{0}_{\mathbf{x}^\ast_{?}}^{(n_?,n_?)}\cdot \mathbf{A}^{j}\cdot \mathbf{b}\right)+\varphi\\
    =~& \iota_{?} \cdot \left(\underbrace{\sum_{j=0}^\infty  \mathbf{A}^j \cdot \mathbf{d}(\mathbf{x}_I)}_{\psi_{I,1}} + \underbrace{\sum_{j=0}^\infty  \mathbf{A}^{j}\cdot\mathbf{C}(\mathbf{x}-I)\cdot\sum_{j=0}^\infty  \mathbf{A}^{j}\cdot \mathbf{b}}_{\psi_{I,2}}\right)+\varphi \enspace,
\end{align*}
for some $\varphi$. We see that $\varphi $ is either $0$ or a polynomial whose least order is not smaller than $2$. Thus,
\begin{align*}
    \lim_{\|\mathbf{x}_I\|\rightarrow 0}\dfrac{\varphi}{\|\mathbf{x}_I\|}=0 \enspace.
\end{align*}
Let $\sum_{i\in I}\mathbf{h}_i\cdot\mathbf{x}_i=\iota_{?}\cdot (\psi_{i,1}+\psi_{i,2})$, and we obtain the equation in Proposition \ref{th:linapproxg}, namely
\begin{align*}
    \varrho(\mathbf{x}_I+\mathbf{r}_I)\approx \iota_{?}\cdot \sum_{i\in I}(\psi_{i,1}+\psi_{i,2})\enspace.
\end{align*}
\end{proof}

\begin{proof}[Proof of Theorem \ref{th:link}]
It holds that for each $i\in I$
\begin{align*}
    \kappa_i = &~\dfrac{1}{2}(\max(\mathbf{h}_i)-\min(\mathbf{h}_i)) \\
    \kappa_w = &~\dfrac{1}{2}\sum_{i\in I} w(i)(\max(\mathbf{h}_i)-\min(\mathbf{h}_i))
\end{align*}
where $w(i)=\Delta_1/\Delta$ and $\Delta=\sum_{i\in I} \Delta_i$. The equation in the theorem follows.
\end{proof}

%%%%%%%%%%%%%%%%%%%%%%%%%%%%%%%%%%%%%%%%%%%%%%%%%%%%%%%%%%%%%%%%%%%%%%%%%%%%%%%%%%%%%%%%%%%%%%%%%%%%%%%%%%%%%%%%%%%%%%%%%%%%%%%%%%%%%%%%%%%%%%%%%%%%%%%%%%%%%%%%%
%%%%%%%%%%%%%%%%%%%%%%%%%%%%%%%%%%%%%%%%%%%%%%%%%%%%%%%%%%%%%%%%%%%%%%%%%%%%%%%%%%%%%%%%%%%%%%%%%%%%%%%%%%%%%%%%%%%%%%%%%%%%%%%%%%%%%%%%%%%%%%%%%%%%%%%%%%%%%%%%%
%%%%%%%%%%%%%%%%%%%%%%%%%%%%%%%%%%%%%%%%%%%%%%%%%%%%%%%%%%%%%%%%%%%%%%%%%%%%%%%%%%%%%%%%%%%%%%%%%%%%%%%%%%%%%%%%%%%%%%%%%%%%%%%%%%%%%%%%%%%%%%%%%%%%%%%%%%%%%%%%%
%%%%%%%%%%%%%%%%%%%%%%%%%%%%%%%%%%%%%%%%%%%%%%%%%%%%%%%%%%%%%%%%%%%%%%%%%%%%%%%%%%%%%%%%%%%%%%%%%%%%%%%%%%%%%%%%%%%%%%%%%%%%%%%%%%%%%%%%%%%%%%%%%%%%%%%%%%%%%%%%%
%%%%%%%%%%%%%%%%%%%%%%%%%%%%%%%%%%%%%%%%%%%%%%%%%%%%%%%%%%%%%%%%%%%%%%%%%%%%%%%%%%%%%%%%%%%%%%%%%%%%%%%%%%%%%%%%%%%%%%%%%%%%%%%%%%%%%%%%%%%%%%%%%%%%%%%%%%%%%%%%%
%%%%%%%%%%%%%%%%%%%%%%%%%%%%%%%%%%%%%%%%%%%%%%%%%%%%%%%%%%%%%%%%%%%%%%%%%%%%%%%%%%%%%%%%%%%%%%%%%%%%%%%%%%%%%%%%%%%%%%%%%%%%%%%%%%%%%%%%%%%%%%%%%%%%%%%%%%%%%%%%%
%%%%%%%%%%%%%%%%%%%%%%%%%%%%%%%%%%%%%%%%%%%%%%%%%%%%%%%%%%%%%%%%%%%%%%%%%%%%%%%%%%%%%%%%%%%%%%%%%%%%%%%%%%%%%%%%%%%%%%%%%%%%%%%%%%%%%%%%%%%%%%%%%%%%%%%%%%%%%%%%%
\end{document}